\def\hatn{{\bf \hat n}}

\def\hatz{{\bf \hat z}}

\def\VEV#1{{\left\langle #1 \right\rangle}}

\newcommand{\be}{\begin{equation}}
\newcommand{\ee}{\end{equation}}
\newcommand{\bea}{\begin{eqnarray}}
\newcommand{\eea}{\end{eqnarray}}
\def\d{\partial}
\def\e{\epsilon}
\def\pslash{{\cal P}{\hbox{\kern-6pt $\slash$}}}

\long\def\comment#1{}

\documentstyle[prl,aps,floats,twocolumn]{revtex}

\input epsf
\begin{document}
\draft
\twocolumn[\hsize\textwidth\columnwidth\hsize\csname @twocolumnfalse\endcsname
\title{Cosmological Signature of New Parity-Violating Interactions}
\author{Arthur Lue,\cite{luemail} Limin Wang,\cite{wangmail} and 
Marc Kamionkowski,\cite{marcmail}}
\address{Department of Physics, Columbia University, 538 West
120th Street, New York, New York 10027}
\date{December 1998}
\maketitle

\begin{abstract}
Does Nature yield any manifestations of parity violation other than
those observed in weak interactions?  A map of the cosmic microwave
background (CMB) temperature and polarization will provide a new
signature of P violation.  We examine two classes of P violating
interactions that would give rise to such a signature.  The first
interaction leads to a cosmological birefringence, possibly
driven by quintessence.  The other interaction leads to to an
asymmetry in the amplitude of right- versus left-handed
gravitational waves produced during inflation.  The Planck Surveyor
should improve upon the current sensitivity to birefringence.
While the primordial effect would most likely elude detection by
MAP and Planck, it may be detectable with a future dedicated CMB 
polarization experiment.
\end{abstract}

\pacs{PACS numbers: 98.80.Es, 04.50.+h  \hfill
CU-TP-926, CAL-675, astro-ph/9812088}
]

The discovery of parity (P) violation \cite{porig}
was central to the development of what has now become the standard
model.  Nevertheless, this symmetry violation occurs strictly within
the weak interaction sector.  Presumably, its ultimate origin lies
in the grand-unified and/or Planck-scale physics that yields the
standard model as its low-energy limit.  If so, might there be some
remnant of P violation in gravitational interactions or in some
other, still undiscovered, sector?

Some tantalizing clues do exist.  The baryon asymmetry of the Universe
requires charge conjugation (C) violation as well as CP violation
\cite{BAU}, likely in new physics beyond the standard model.
Moreover, extensions of the standard model, including grand unified
theories and supersymmetry, naturally suggest nonstandard P and CP
violating interactions.  Carroll has argued that a certain class of
quintessence models should generically produce such P asymmetric
physics (``cosmological birefringence'')
\cite{Carroll}, and other cosmological physics may also give
rise to parity breaking \cite{CF}.

In the next few years, high-precision temperature and polarization
maps of the cosmic microwave background (CMB) will become available
\cite{MAP,Planck}.  These maps will provide a wealth of data
concerning the physics of the early Universe.  Although the primary
purpose of these observations is not to explore P violation,
certain temperature/polarization cross-correlation functions can
provide a probe of P violation.  Their relevance has heretofore
been disregarded since they vanish if the underlying physics---in
particular gravity and inflation---is P symmetric, as has been
assumed until now.

In this Letter we explore the possibility of probing exotic
P violating physics using the CMB.  We first lay out the details
of the CMB correlation functions needed to detect P
violation.  We then explain the features of
fundamental interactions and early-Universe mechanisms required to
produce such a preferred macroscopic orientation.  We then provide two
examples of interactions and mechanisms that can produce this
P violating signature and discuss their detectability.

A map of the temperature $T(\hatn)$ as a function of position
$\hatn$ on the sky can be expanded in spherical harmonics,
$Y_{(lm)}$, with expansion coefficients $a^T_{(lm)}$
given by the inverse transformation that follows from the
orthonormality of the spherical harmonics.  
Suppose that in addition, the Stokes parameters $Q(\hatn)$ and
$U(\hatn)$ required to specify the linear-polarization state are
also mapped.  The Stokes parameters are components of a $2\times
2$ symmetric trace-free tensor.  As detailed in
Refs. \cite{KKS,zs}, this polarization tensor field can be
expanded in tensor spherical harmonics $Y_{(lm)ab}^G(\hatn)$
and $Y_{(lm)ab}^C(\hatn)$, which are a complete basis for the
``gradient'' (i.e., curl-free) and ``curl'' components of the
tensor field, respectively.  The expansion coefficients
$a_{(lm)}^G$ and $a_{(lm)}^C$ for the gradient and curl
components, respectively, can be obtained from the inverse
transformations that follow from the orthonormality properties
of these  tensor harmonics.

The $a_{(lm)}^X$'s (for $X=\{T,G,C\}$) have zero mean
$\VEV{a_{(lm)}^X}=0$ and covariances $\VEV{a_{(lm)}^X
(a_{(lm)}^{X'})^*} = C_l^{XX'}$, when averaged over an ensemble of
Universes.  For the single Universe that we observe, each $C_l^{XX'}$
can be estimated from the $2l+1$ individual $m$ modes.  The two-point
statistics of the temperature/polarization map are thus completely
specified by the six ($TT$, $GG$, $CC$, $TG$, $TC$, and $GC$) sets of
multipole moments.  If the temperature/polarization distribution is P
invariant, then $C_l^{TC}$ and $C_l^{GC}$ must vanish because the
$Y_{(lm)}$ and the $Y_{(lm)ab}^{G}$ have parity $(-1)^l$ while the
$Y_{(lm)ab}^{C}$ have parity $(-1)^{l+1}$
\cite{kksprl,szprl}.  Therefore, if $C_l^{TC}$ and/or $C_l^{GC}$ 
is found to be nonzero with some statistical significance, it
indicates a preferred orientation in our Universe.

What physics would be required to produce such a P violating CMB
temperature/polarization pattern?  This P violation is different from
that in weak interactions since weak interactions are P violating only
if the particle-antiparticle character is known; they
would be P conserving in an experiment which did not discriminate
between particles and antiparticles (neglecting the small CP violation
in the standard model).  This CMB signature is charge-blind: it
requires a preferred handedness.

The existence of interactions that yield P asymmetric physics alone
are insufficient to produce a preferred cosmological orientation from
a P symmetric initial state.  Since the CMB signature is charge-blind,
then the CPT theorem suggests that the required interaction must
violate time-reversal (T) invariance as well as P invariance in a
fashion that preserves PT.  If we have an interaction that is P and T
violating, then any mechanism that defines an arrow of time could
conceivably drive the Universe to a preferred orientation.  Such a T
asymmetric process might be the expansion of the Universe or maybe
some entropy-producing process.  Another possibility, and that which
we focus on here, is that the T symmetry is broken by the rolling of
some scalar field.

If there is some P and T violating physics that appears at some
large energy scale $\mu$ that involves a new scalar field
$\chi$, then at lower energies we would expect terms in our
effective Lagrangian like
\be
    {\cal L}_{int} = g(\chi) F_{\mu \nu}\tilde{F}^{\mu \nu},
\label{eq:CFterm}
\ee
where $g(\chi)$ is a dimensionless function of a scalar field
and $F_{\mu\nu}$ is the electromagnetic field-strength tensor.
The scalar field $\chi$ has been identified, e.g., with that in
scalar-tensor theories of gravity \cite{CF} or with a quintessence
field \cite{Carroll}.  If $\chi$ is constant in space and time, then
the term has no effect on electrodynamics, since the term can be
written as a total derivative.  However, if $\chi$ is spatially
homogeneous but changing with time, then the polarization vector of a
photon is rotated by an angle $\Delta \alpha \propto \Delta g(\chi)$,
where $\Delta g(\chi)$ is the change in the function $g(\chi)$ as the
photon propagates from source to observer \cite{CF}; this
effect has been referred to as ``cosmological birefringence.''

\begin{figure}[t]  
\epsfxsize=3.3 in \epsfbox{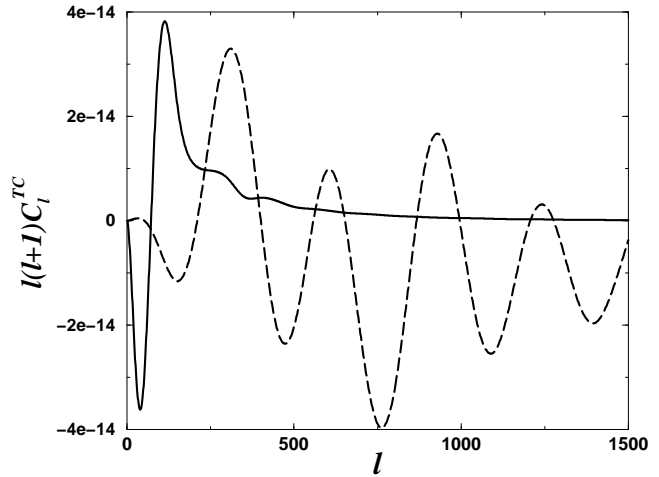}  
\caption{The dashed curve shows the $C_l^{TC}$ power spectrum
induced by rotation of the polarization of an initially P
symmetric CMB polarization pattern by $0.05^\circ$.  The solid
curve shows the $C_l^{TC}$ power spectrum produced by a
GW background that consists of only right-handed GWs.
}
\label{fig:sensi}
\end{figure}

The effect of such a rotation is to alter a P-symmetric CMB as
it propagates from the surface of last scatter to the observer.  
Each $Y_{(lm)}^G$ tensor field is orthogonal to the
$Y_{(lm)}^C$ of the same $l$ and $m$ at each point on the sky,
so rotating the polarization of each photon everywhere by the same amount
simply mixes the $G$ and $C$ modes.  Any mechanism that produces
temperature anisotropies also produces a polarization
pattern with a gradient component, and it also produces a
non-zero $TG$ cross-correlation.  If the CMB has some
nonzero $C_l^{TG}$ moments at the surface of last scatter, and
the polarization vector of each photon is rotated by an angle
$\Delta\alpha$, then it induces $TC$ moments, $C_l^{TC}
= C_l^{TG} \sin{2\Delta\alpha}$.
Furthermore, the shape of the $C_l^{TC}$ power spectrum (as a
function of $l$) is the same as that of the $C_l^{TG}$
power spectrum.  The dashed curve in  Fig. \ref{fig:sensi} shows 
an example of such a $C_l^{TC}$ power spectrum.  This curve was
generated assuming a flat model with a matter density
$\Omega_m=0.3$, a cosmological constant $\Omega_\Lambda = 0.7$,
a baryon density $\Omega_bh^2 = 0.02$, and Hubble
parameter $h=0.65$ with a nearly scale-invariant spectrum of
primordial adiabatic perturbations and no gravitational waves.

Let us now consider the consequences of another class of terms that
generically appears in our effective action
\be
	{\cal L}_{int} = 
	f(\Phi)R^\lambda_{\ \sigma\mu\nu}\tilde{R}_\lambda^{\ \sigma\mu\nu}\ .
\label{spslash}
\ee
In contrast to our earlier discussion, here we identify scalar field
$\Phi$ with the inflaton field.  These terms arise in exact analogy
from whatever physics that produces terms like Eq. (\ref{eq:CFterm}).
For example, let $\Phi$ be an axion- or pion-like field axially
coupled to heavy fermions.  Then, radiative fermion loops generate both
Eqs. (\ref{eq:CFterm}) and (\ref{spslash}).  Another class of
examples appears in \cite{new}.

So long as the scalar field is homogeneous and constant in time, Eq.
(\ref{spslash}) becomes a pure surface term, and thus does not
contribute at all to classical gravity dynamics.  Thus we expect that
after inflation, when the inflaton has come to rest, P asymmetric
gravity dynamics is not present, suggesting no current observed
constraint on Eq. (\ref{spslash}).  Nevertheless, the term has
relevant effects during inflation which may be observed through the
CMB.

The homogeneous dynamics of the inflaton is identical to that without
Eq. (\ref{spslash}).  We may take any conventional slow-roll
inflation scenario where $\dot{\Phi} \neq 0$.  Moreover, the
conventional flat Robertson-Walker metric is still a solution to the
metric equations of motion with the new interactions, implying the
overall cosmology is not affected by the new term.  However, metric
perturbations are affected by these terms.  For simplicity, take
the metric $g_{\alpha\beta} =
\eta_{\alpha\beta} + h_{\alpha\beta}$ in a flat-space background,
$\eta_{\alpha\beta}$.  Linearizing the metric equations of motion in
the harmonic gauge
($\d^\nu h_{\mu\nu} = \frac{1}{2}\d_\mu h^\nu_\nu$), we find
\bea
M_P^2\Box h_{\alpha\beta} &=& 2f^{\prime\prime}\dot{\Phi}^2\e^{ijk}
\eta_{i\alpha}(\d_\beta\d_jh_{0k}+\d_0\d_kh_{\beta j})	\\
&\ & + 2f' \dot{\Phi}\e^{ijk}\eta_{i\alpha}\d_j\Box h_{\beta k} +
(\alpha\leftrightarrow\beta), \nonumber
\eea
assuming the acceleration of the inflaton is negligible.  Here, the
prime on $f$ denotes differentiation with respect to $\Phi$, and the
Latin indices indicate spatial indices only.  Let us look at
plane-wave perturbations of the form $h_{\alpha\beta} =
e_{\alpha\beta} e^{-ik\cdot x}$ where $e_{\alpha\beta}$ is a constant
polarization matrix. Assuming the effects of the new terms are small,
we find the following plane-wave solutions
\be
	e^R_{\mu\nu}e^{2f^{\prime\prime}
	\dot{\Phi}^2kt/M_P^2}e^{-ikt+ikz},
	e^L_{\mu\nu}e^{-2f^{\prime\prime}
	\dot{\Phi}^2kt/M_P^2}e^{-ikt+ikz}
\ee
where $e^R,e^L$ are the polarization tensors for right- and
left-handed polarized waves, respectively.  Thus, 
right-handed gravitational waves (GWs) are amplified as they propagate while
left-handed GWs are attenuated.  These solutions preserve PT
although they violate P and T individually.

Let us apply this result to our scenario where the Universe inflates.
While their wavelength is much smaller than the horizon size,
right-handed GWs amplify while left-handed GWs attenuate.
Eventually, the fluctuations expand past the horizon and freeze
out.  To estimate the discrepancy between left-handed and right-handed
tensor fluctuations in the early Universe, we assume that the fluctuations
of both handednesses are equal in amplitude and behave classically as
they expand beyond a wavelength $1/\mu$ and then freeze as the
wavelength becomes comparable to the horizon scale.  When the waves
exit the horizon scale, we can estimate the fraction of accumulated
discrepancy through the index $\e$:
\be
	\e \sim (M_P/\mu)
		(H/ M_P)^3(\dot{\Phi}/ H^2)^2,
\ee
where $H$ is the Hubble scale and $f^{\prime\prime}$ is characterized
by the scale $1/\mu^2$.  The factor $H^2/\dot{\Phi}$ is
associated with the amplitude of scalar density perturbations ($\sim
10^{-5}$), while the factor $H/M_P$ is associated with the amplitude
of tensor perturbations ($<3\times 10^{-6}$).  Given fixed
cosmological parameters, one may think of a limit on $\e$ as
a lower bound on $\mu/M_P$.

Let us describe how this physics is reflected in the CMB.
Long-wavelength GWs produce temperature anisotropies and also a curl
component of the polarization
\cite{kksprl,szprl}.  An excess of right over left (or {\it vice
versa}) circularly-polarized GWs produces a nonzero $C_l^{TC}$.
Consider a single right-handed circularly polarized GW with wavenumber
$k$ propagating in the $+\hatz$ direction. This can be written as an
out-of-phase combination of a linearly-polarized GW with $+$
polarization and another of equal amplitude with a $\times$
polarization.  We can always choose the $x$ and $y$ axes so that the
amplitude of the $+$ component has a crest at the origin and the
$\times$ component has a zero at the origin.  Doing so, the multipole
coefficients $a_{(lm)}^T$ for the temperature pattern induced on the
sky by this particular circularly-polarized wave is
\cite{arthur}
\be
     a_{(lm)}^T = \cases{ (\delta_{m,2}+ \delta_{m,-2}) A_l^T(k)
     & even $l$ ($+$), \cr
     -i (\delta_{m,2}-\delta_{m,-2}) A_l^T(k) & odd $l$ ($\times$), \cr }
\label{eq:Tmoments}
\ee
where, as indicated, the even-$l$ contribution is from the $+$
mode and the odd-$l$ contribution is from the $\times$ mode.
For a left-handed circularly-polarized wave, the sign of the
odd-$l$ moments (from the $\times$ contribution) are reversed.
The $A_l^T(k)$ are temperature brightness functions (see
\cite{KKS}).  The multipole coefficients for the $G$ component
of the CMB polarization are similar, except that the $A_l^T(k)$ are
replaced by some polarization functions $A_l^G(k)$.  The
multipole coefficients for the C component of the CMB
polarization is similar,
\be
     a_{(lm)}^C = \cases{ (\delta_{m,2}+ \delta_{m,-2}) A_l^C(k)
     & even $l$ ($\times$), \cr
     -i (\delta_{m,2}-\delta_{m,-2}) A_l^C(k) & odd $l$ ($+$), \cr }
\label{eq:Cmoments}
\ee
except note that the even-$l$ moments now come from the $\times$ 
mode and the odd-$l$ moments come from the $+$ mode.  For 
a left-handed wave the sign of the even-$l$ moments is
reversed.

Eqs.~(\ref{eq:Tmoments}) and (\ref{eq:Cmoments}) indicate why
$C_l^{TC}=0$ (and why $C_l^{GC}=0$) for linearly-polarized
waves.  For example, if we have only a $+$ polarized wave, then
the T pattern induces only even-$l$ modes and the C pattern
induces only odd-$l$ modes.  But these equations also show that
a circularly-polarized wave induces a nonzero $C_l^{TC}$.
Recall that we measure a given $C_l^{TC}$ by averaging the
quantity $a_{(lm)}^T (a_{(lm)}^C)^*$ over all $2l+1$ values of
$m$.  Doing so, we find that this right-handed GW
induces a nonvanishing $C_l^{TC}=2(2l+1)^{-1} A_l^T(k)
A_l^C(k)$, and a left-handed GW induces the same 
quantity but with the opposite sign.  Since $C_l^{TC}$ is
rotationally invariant, the result is independent of the
direction of propagation of the GW.

\begin{figure}[t]  
\epsfxsize=3.3 in \epsfbox{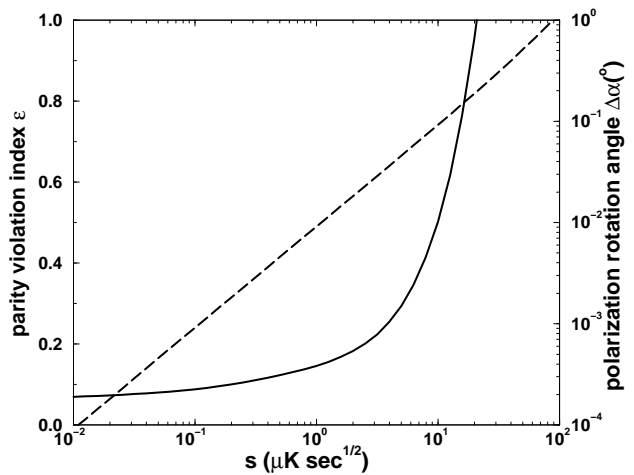}  
\caption{The smallest $\e$ for the (GW model)
     and $\Delta\alpha$ (for the polarization rotation model)
     detectable at the $1\sigma$ level with a one-year CMB
     temperature/polarization experiment with
     detector sensitivity $s$.  For this calculation, a beamwidth of
     $0.1^{\circ}$ is assumed (although results
     for $\e$ are roughly the same for a beamwidth as large as
     $0.5^\circ$).}
\label{fig:sensit}
\end{figure}

The solid curve in Fig.~\ref{fig:sensi} shows $C_l^{TC,R}$, the
$TC$ power spectrum expected for a GW background
made of only right-handed GWs.  This curve was generated
assuming the same classical cosmological parameters as were used 
for the dashed curve, but here we have assumed the presence of a
nearly scale-invariant spectrum of GWs with a
tensor-to-scalar ratio of ${\cal T}/{\cal S}=0.7$.
For a more general mixture of
right- and left-handed GWs, $C_l^{TC} = \e C_l^{TC,R}$.
The solid curve in Fig.~\ref{fig:sensit} shows the smallest $\e$ 
that could be distinguished from a null result from the
$C_l^{TC}$ moments at the $1\sigma$
level as a function of detector sensitivity $s$ for a one-year
experiment that maps the temperature and polarization of the
entire sky.  The calculation was done using the same
cosmological parameters as were used in Fig. \ref{fig:sensi}.

The sensitivity to $\e$ remains finite even as $s\rightarrow0$, since
the measurement is ultimately cosmic-variance limited.
Fig. \ref{fig:sensi} is only meant to be illustrative; the precise
sensitivity differs for different cosmological
parameters.  
The sensitivity to $\e$ will of course be degraded if the
tensor amplitude is smaller.  Because $\e < 1$ is a strict constraint,
neither MAP ($s\simeq150~\mu {\rm K}~\sqrt{\rm sec}$)
\cite{MAP} nor Planck ($s\simeq35~\mu {\rm K}~\sqrt{\rm sec}$)
\cite{Planck} will be able to detect any such left-right asymmetry, but
a post-Planck experiment might conceivably be able to discriminate a
value as small as $\e \sim 0.08$.  With the cosmological parameters used,
this value of $\e$
corresponds to having th P and T violating physics occur at the scale
$\mu \sim 4\times 10^{-5} M_P$.  This discussion in some sense
conservative since we have not considered the additional information
provided by the $C_l^{GC}$ moments or the improved sensitivity
possible with a deeper map of a smaller region of sky.

Similarly, the dashed curve in Fig.~\ref{fig:sensit} shows the
smallest rotation angle $\Delta\alpha$ induced by the term in
Eq. (\ref{eq:CFterm}) that could be distinguished from a null result
at the 1$\sigma$ level as a function of $s$.  Again, the underlying
cosmological parameters are taken to be those used in
Fig. \ref{fig:sensi}.  Note that with no tensor perturbations, there
is no cosmic-variance limit to the detectability of $\Delta\alpha$.
Correlations between the elongation axes and polarization vectors of
distant radio galaxies and quasars can put constraints on $\Delta
\alpha$ at the order of $1^{\circ}$ \cite{Carroll}.
Figure~\ref{fig:sensit} shows that the Planck Surveyor is slightly
more sensitive, while a future high-precision CMB polarization could
provide much better sensitivity, e.g., $\Delta\alpha \lesssim
0.01^{\circ}$ for $s \lesssim 1~\mu {\rm K}~\sqrt{\rm sec}$.
Moreover, radio sources only probe the motion of the scalar field
between now and redshifts of a few, whereas the CMB probes the motion
of the scalar field out to redshifts $z\simeq1100$.  Thus, the CMB
should provide a better probe of models such as quintessence models
with a tracking solution \cite{quint}, in which the scalar field is
expected to do most of its rolling at early times.

Non-zero $C_l^{TC}$ can similarly be induced by Faraday rotation 
due to intervening magnetic fields \cite{ferreira}.  
However, Faraday rotation depends
on the CMB photon frequency \cite{arthuravi}, whereas the effects
we are considering are frequency independent. Furthermore,
Faraday rotation is an anisotropic effect, so it affects the
the $l$-dependence of the $C_l^{TC}$ (unless the magnetic field
is very homogeneous in which case only the very lowest-$l$ modes
would be affected).

Should inflationary or quintessence physics be P and T violating,
these effects should in general be present, and if detected, would
provide a valuable window to cosmological physics.  There may be other
sources of parity breaking in addition to those we discuss that would
engender the CMB signature considered.  A dedicated CMB polarization
experiment would be poised to yield a wealth of new information about
the early Universe.  We have shown here that such observations would
also be capable of providing unique tests of exotic P violation.

\medskip
We used a modified version of CMBFAST \cite{zs,edbert} to calculate
the $C_l^{TC}$.  We thank E. Weinberg, B. Greene, and S. Carroll for
useful discussions.  This work was supported by a DoE Outstanding
Junior Investigator Award, DE-FG02-92ER40699, NASA grant NAG5-3091,
and the Alfred P. Sloan Foundation.

\end{document}